\documentclass[namedreferences]{solarphysics}
\usepackage[hyperref,optionalrh,solaromanenum]{spr-sola-addons} 
\usepackage{graphicx}                    
\usepackage{amssymb}                    
\usepackage{color}                       
\usepackage{breakurl}                         

\begin{document}

\begin{article}

\begin{opening}

\title{The latitudinal migration of the sunspot groups}

%
\author[addressref={aff1},corref,email={jouni.j.takalo@oulu.fi}]{\inits{J.J.}\fnm{Jouni}~\lnm{Takalo}}

\institute{$^{1}$ Space physics and astronomy research unit, University of Oulu, POB 3000, FIN-90014, Oulu, Finland\\
\email{jouni.j.takalo@oulu.fi}
}
%
\runningauthor{J. Takalo}
\runningtitle{Drift of sunspot groups}


\begin{abstract}
We  analyse the monthly sunspot group (SG) data for Solar Cycles C8\,--\,C23 and calculate the average latitude of the drift path for the northern and southern hemisphere of the Sun. We find that exponential function fits slightly better than the second-order polynomial to the average drift of the SGs. The drift velocities are are 0.30 and 0.27 degrees/month (meridional speed $\approx$\,1.41 and 1.27 m/s) in the beginning and 0.071 and 0.069 degrees/month ($\approx$\,0.33 and 0.32 m/s) at the end of the average cycle for northern and southern hemisphere, respectively.

We do also bi-linear fits (first-order polynomial fits in two fractions), because they tell us the crossing point were the drift velocity changes the most. This crossing point for the Cycles C8\--\,C23 is at 50 months and 15.4 degrees of latitude for the northern hemisphere and 54 months and -14.4 degrees for the southern hemisphere. The drift velocities calculated from the slopes of the bi-linear fits are 0.233 and 0.216 degrees/month ($\approx$\,1.10 and 1.02 m/s) before the crossing point and 0.118 and 0.101 degrees/month ($\approx$\,0.55 and 0.47 m/s) after the crossing point.

We find a slight dependence of the migration on the strength of the cycle such that the drift path of the stronger cycles follow slightly higher latitudes than those of the weaker cycles. The areas of the sunspot groups affect also slightly to the drift path such that large area SGs have path at somewhat lower latitudes in the ascending phase and higher latitudes in the descending phase of the cycle than the smaller area SGs.
 
\end{abstract}

%

\end{opening}


\section{Introduction}

The most striking pattern in solar research was discovered by Maunder in studying the occurrence of the sunspots 1874\,--\,1902 \citep{Maunder_1904}. Especially, when plotting the distribution of the sunspots during 1879\,--\,1889 (Solar Cycle 12) he noticed how the sunspots systemically approached the Equator of the Sun during the course of the years. S.Carrington had noted already in the transition of the Cycle 9 to Cycle 10 that sunspots disappeared from the low latitudes and soon appeared at latitudes between 20\,--\,40 degrees of heliographic latitudes. He reported this phenomenon in his article \citep{Carrington_1858}. The shape of the diagram, resembling butterfly, was first drawn by Maunder for the Cycle 12 and 13, but he did not use that name in his original paper \citep{Maunder_1904}. \cite{Gleissberg_1971} mention that it is not known who used the name "`Butterfly diagram"' for the first time, but at least Maunder used the name by 1922. In this same article \cite{Gleissberg_1971} find that the mean latitude of sunspots decrease  0.012\,$\times\,sin(2\phi_{m})$ degrees/day, where $\phi_{m}$ is the heliographic latitude..

Later there have been some additional studies to find the formula for the average drift and velocity of the migration for the sunspot and sunspot groups.
\cite{Li_2001} found the mean latitude for the Royal Greenwich Observatory sunspot group data for Cycles 12\,--\,22. They used quarterly year data and fitted second-order polynomial to the overlapped data. The equations of the fitting curves were $l$=0.0893$\times t^{2}$-2.8$\times$t+27.24 and $l$=-0.0767$\times t^{2}$+2.6$\times$t-26.72 and correlation coefficients 0.9753 and 0.9749 for the fits of the northern and southern hemisphere, respectively. They state that the migration velocity is largest in the beginning of the cycle and the latitude decreases with an average of 1.6 degrees and 1.56 degrees in the northern and southern hemisphere, respectively.

\cite{Hathaway_2011} used the same data but for the Cycles 12\,--\,23 in his analysis. He stated that there is no dependence of the migration path on the strength of the cycle and found a standard exponential equation for the trajectory of the mean latitude, i.e. $\overline{\lambda} = 28^{o}\, exp(-(t\,-\,t_{0})/90)$
, where t is measured in months.

\cite{Zhang_2018} studied the Extended time series of Solar Activity Indices (ESAI) for the years 1854-1985. They used yearly averaged data and were interested if the migration of sunspots is linear or nonlinear. They found that second-order polynomial fit outshines the linear (first-order) fit in their analysis. They found that migration velocity is about 2.2 degrees/year in the start of the cycle and decreases to 1.8 degrees/year in the maximum phase and is 0.7 degrees/year at the end of the average cycle.

In this study we analyse the average drift of the sunspot groups for Solar Cycles (SC) 8\,--\,23 from three different databases and find the best-fit functions for the evolution of the drift. This article is organized as follows. Section 2 presents the used data. In Section 3 we study and discuss different scenarios of the migration and give our conclusion in Section 4.

\section{Data and Methods}

\subsection{Sunspot Group Data}

For the analyses of solar cycles we use two different sunspot group datasets used also earlier by \cite{Takalo_2020_1, Takalo_2020_2}. The newest one is recently published catalogue of sunspot groups by \cite{Mandal_2020} (hereafter M set). The main part of this dataset consists of data from the Royal Greenwich observatory (RGO) measurements between 1874-1976. In this time series the composite data sets from Kislovodsk, Pulkova and Debrecen observatories were used for the years from 1976 to October 2019, i.e., containing (almost) solar cycle 24. We, however, omit the Cycle 24 here, because it is incomplete and study Cycles C12\,--\,23. The other dataset, which we use is the sunspot group data for SC8\,--\,SC23 by \cite{Leussu_2017a, Leussu_2017b} (hereafter L set). The years 1874\,--\,1976 for this data are also from the RGO dataset, but the solar cycles SC21\,--\,SC23  are constructed from SOON (Solar Observatory Optical Network) data for the period 1976-2015 \citep{Neidig_1998}. Again the Cycle 24 is incomplete, and we do not include it to the analyses. This dataset has been  completed using digitized observations of S. H. Schwabe \citep {Arlt_2011, Arlt_2013}, which provide an extension of the data to the years 1825\,--\,–1867, covering total Solar Cycles 8\,--\,10. Later this dataset has been filled with Cycle 11 using newly digitized Sp\"orer data \citep{Diercke_2015} for the years 1868\,--\,–1874 \citep{Leussu_2017a}.
There are differences in the datasets such that the L set does not contain the area data of the sunspot groups, while M set includes also the areas. In addition, the number of groups in L set is much smaller, because it contains only the first appearance of the sunspot group in its record, while M set contains daily appearance of the groups. Thus the lifetime of each group also affects the number of the groups in M set.
When comparing the sunspot group data to separate sunspot data, we use also recently published Kodaikanal dataset for the Cycles C16\,--\,23 (hereafter K set) \citep{Jha_2019, Jha_2022}. The minima and lengths for the Solar Cycles 12\,--\,23 used in this study are listed in Table 1. Note, that the L set database already contains northern and southern wings for separate cycles.

\begin{table}
\small
\caption{Sunspot-cycle lengths and dates [fractional years, and year and month] of (starting) sunspot minima for Solar Cycles 12\,--\,23 \citep{NGDC_2013}.}
\begin{tabular}{ c  c  l  c }
  Sunspot cycle    &Fractional    &Year and month     &Cycle length  \\
      number    &year of minimum   & of minimum     &    [years] \\
\hline
12    & 1879.0  &1878 December & 10.6 \\				
13    & 1889.6  &1889 August & 12.1 \\			
14    & 1901.7  &1901 September  & 11.8  \\
15    & 1913.5  &1913 July  &  10.1  \\
16    & 1923.6  &1923 August  & 10.1  \\			
17    & 1933.7  &1933 September & 10.4 \\			
18    & 1944.1  &1944 February  & 10.2  \\
19    & 1954.3  &1954 April  & 10.5  \\
20    & 1964.8  &1964 October  & 11.7  \\
21    & 1976.5  &1976 June & 10.2  \\
22    & 1986.7  &1986 September  & 10.1  \\
23    & 1996.8  &1996 October  & 12.2  \\
24    & 2009.0  &2008 December  

\end{tabular}
\end{table}

\subsection{Resampled Cycles}

To be able to add up cycles with different lengths in order to get more data for solid statistics, we resample the monthly sunspot group (SG) and sunspot data such that all cycles have the same length of 132 time steps (months), i.e. 11 years. This is near the average length of the Cycles 8\,--\,23.

\section{Results and Discussion}

\begin{figure}
	\centering
	\includegraphics[width=1\textwidth]{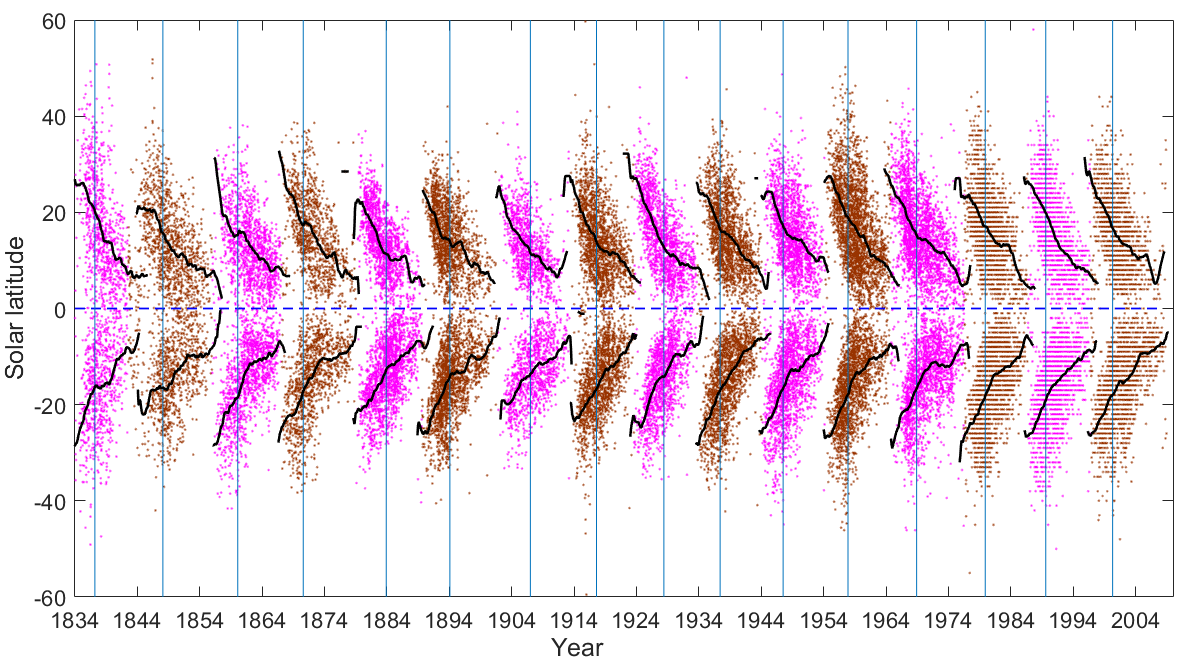}
		\caption{Butterfly pattern of the sunspot groups for L set used in the analysis. The black curves are averages of the sunspot groups calculated for the set. The blue vertical lines are the maxima of the cycles.}
		\label{fig:Leussu_data}
\end{figure}

We start with the analysis with the Lset, which is the longest dataset and gives thus the most solid statistics. Figure \ref{fig:Leussu_data} shows the pattern of the butterfly diagram of the whole L set. The black curves are scarce averages of the sunspot groups calculated for the L set. Note, that in this set the Cycles 21\,--\,23 have only integer values as latitudes. It is, however, evident that the drift of the sunspot groups during the evolution of the cycle is not linear.

\begin{figure}
	\centering
	\includegraphics[width=1.0\textwidth]{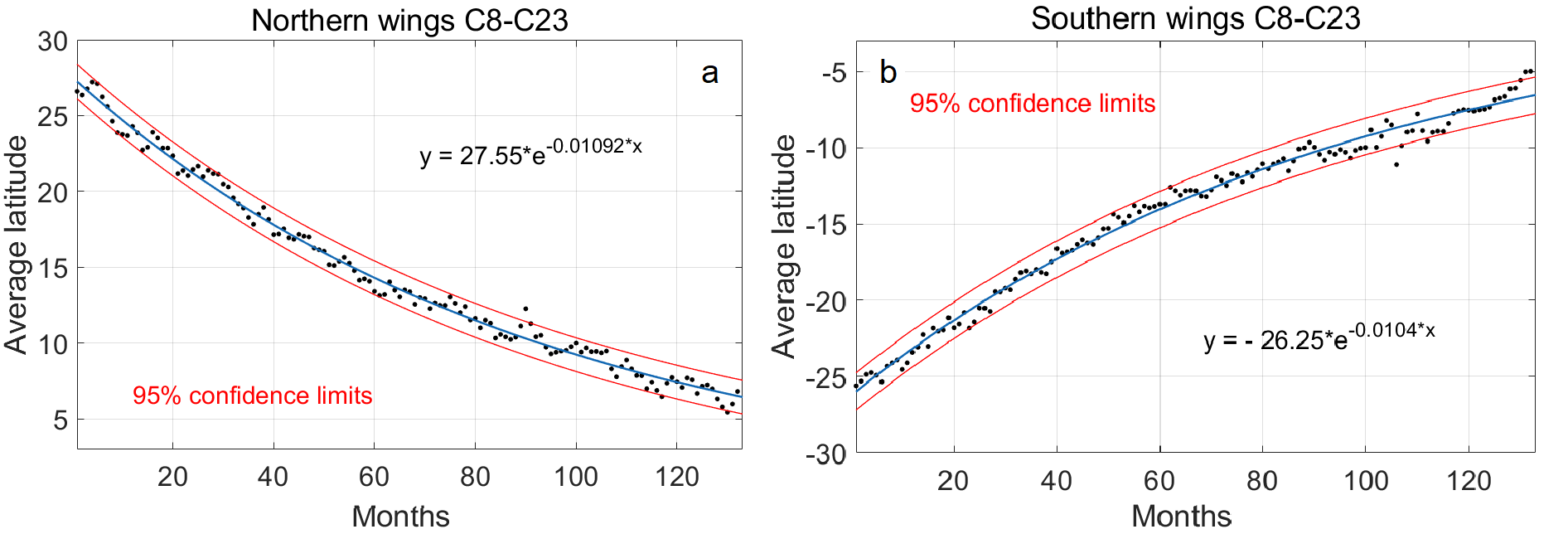}
		\caption{a) The latitudinal drift of the average sunspot group of the northern hemisphere for the L set (Cycles 8\,--\,23) with exponential fit. b) Same as a) but for the southern hemisphere.}
		\label{fig:Lset_NH_SH_average_latitude}
\end{figure}

In order to study the drift of the average latitude of the SGs in more detail, we plot the monthly latitudes of the average SGs in Figure \ref{fig:Lset_NH_SH_average_latitude}a and b for northern and southern hemisphere (NH and SH wings) of the L set, respectively. We made three different fits for the points of the average latitudes; exponential, bi-linear (two first order polynomials) and second order polynomial. The exponential fits are slightly better than the second order fits. The R$^{2}$s are 0.992 and 0.988 for exponential fits of northern and southern hemispheres, respectively. The second order polynomial fits have R$^{2}$s 0.991 and 0.987 for northern and southern hemispheres, respectively. In Figure \ref{fig:Lset_NH_SH_average_latitude} we show the exponential fits with 95\,\% confidence limits. The equations of the exponential fits are

\begin{equation}
	\overline{\lambda} = 27.55\times e^{-0.01092\times t}
\end{equation}
and
\begin{equation}
	\overline{\lambda} = -26.25\times e^{-0.0104\times t}
\end{equation}
for the northern and southern hemisphere, respectively. Here $\overline{\lambda}$ is the estimated average latitude of the SGs and t is time in months.  Note, that we do not make any area weighting, because L dataset does not contain area information. 

\begin{figure}
	\centering
	\includegraphics[width=1.0\textwidth]{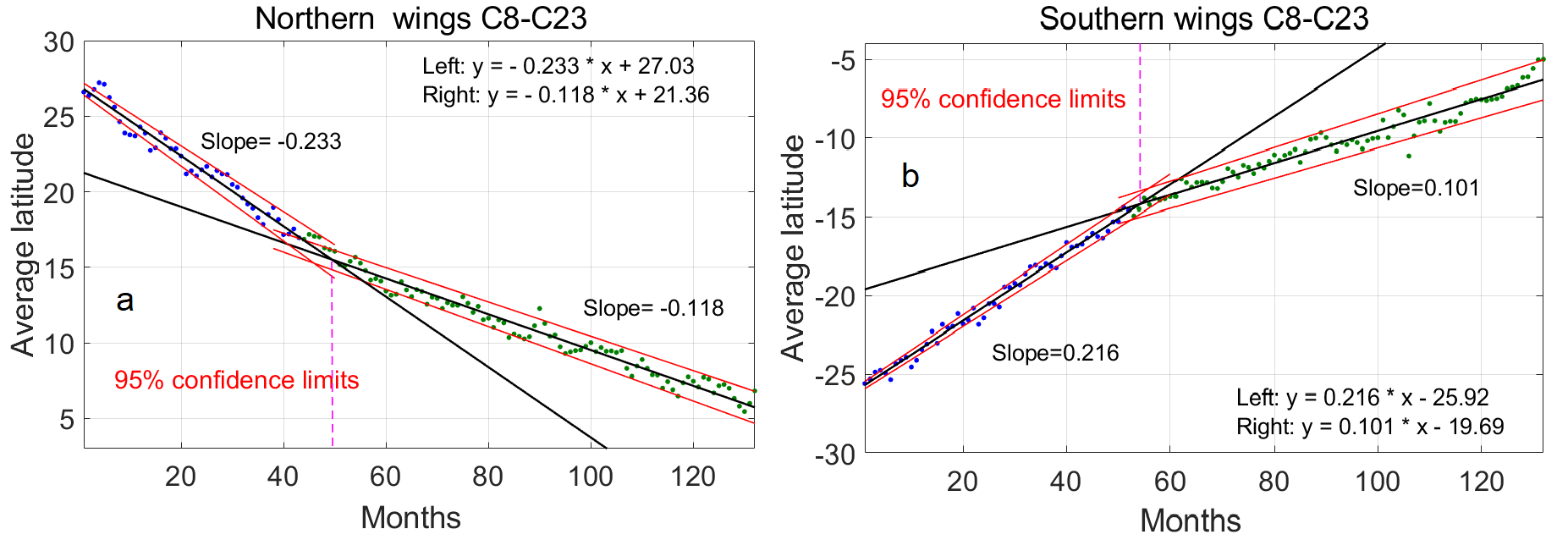}
		\caption{Same data points as in the Figure 2 but now with bi-linear fits. a) Northern
hemisphere b) Southern hemisphere.}
		\label{fig:Lset_NH_SH_Linear_fit}
\end{figure}

\begin{figure}
	\centering
	\includegraphics[width=0.7\textwidth]{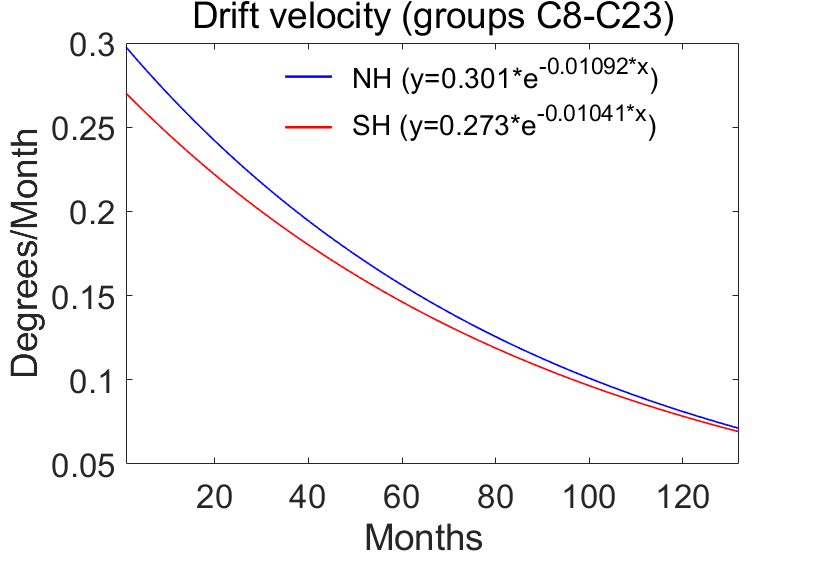}
		\caption{The momentary drift velocities of the average SGs on the northern and southern hemisphere of the Sun.}
		\label{fig:Lset_drift:velocities}
\end{figure}

Figures \ref{fig:Lset_NH_SH_Linear_fit}a and b show the same data point as Figures \ref{fig:Lset_NH_SH_average_latitude}a and b, but now with linear fits which is made in two fragments (bi-linear fit). We have also determined the knee (crossing point) such that the difference of the slopes is largest. The equation of the lines are also shown in the figures with the lines of the 95\,\% confidence limits. Notice, that the confidence limits are in this case narrower towards the higher latitudes. The overall R$^{2}$s are 0.961 and 0.966 for the first and second parts of the northern hemisphere fits and 0.986 and 0.944 for the first and second parts of the southern hemisphere fits, respectively. However, the advantage of the bi-linear fit is that we can estimate the crossing point where the slope radically changes. According to our earlier studies we assumed this to be near the latitude of 15 degrees, and it indeed turned out to be near of that, i.e. the coordinates of the crossing points are (50 months, 15.4 degrees) and (54 months, -14.4 degrees) for the northern and southern hemisphere, respectively. The so-called Gnevyshev gap is also located at the time when the mean latitude of the sunspots crosses about latitude of 15 degrees and this point is kind of a separatrix that divides the solar cycle into two parts, the ascending phase and the declining phase \citep{Takalo_2018, Takalo_2020_1, Takalo_2020_2}.The drift velocities can be estimated the slopes of the fitting lines, i.e 0.233 and 0.216 degrees/month (meridional speed$\approx$\,1.10 and 1.02 m/s) before the crossing point (ascending phase of the cycle) and 0.118 and 0.101 degrees/month ($\approx$\,0.55 and 0.47 m/s) after the crossing point (descending phase of the cycle).
On the other hand, the advantage of the exponential fit is, that we can calculate the drift velocity exactly (point-by-point) as a derivative of the equation of the exponential fit. Figure \ref{fig:Lset_drift:velocities} shows the average momentary drift velocities. The equations for the velocities are $v_{\overline{\lambda}}=0.301\times exp(-0.01092\times t)$ and $v_{\overline{\lambda}}=0.273\times exp(-0.01041\times t)$ (where t is in months) for the northern and southern hemisphere, respectively.The numerical values for the velocities are 0.30 and 0.27 degrees/month ($\approx$\,1.41 and 1.27 m/s) and  at the beginning and 0.071 and 0.069 degrees/month ($\approx$\,0.33 and 0.32 m/s) at the end of the average cycle for northern and southern hemisphere, respectively. Note, that the velocity curves of exponential fits are also exponential. On  the other hand, in the second-order polynomial fits the velocity curves would be linear. Because exponential fits turned to be slightly better than other fits, we use mainly this fitting method in the rest of this study.

\begin{figure}
	\centering
	\includegraphics[width=1\textwidth]{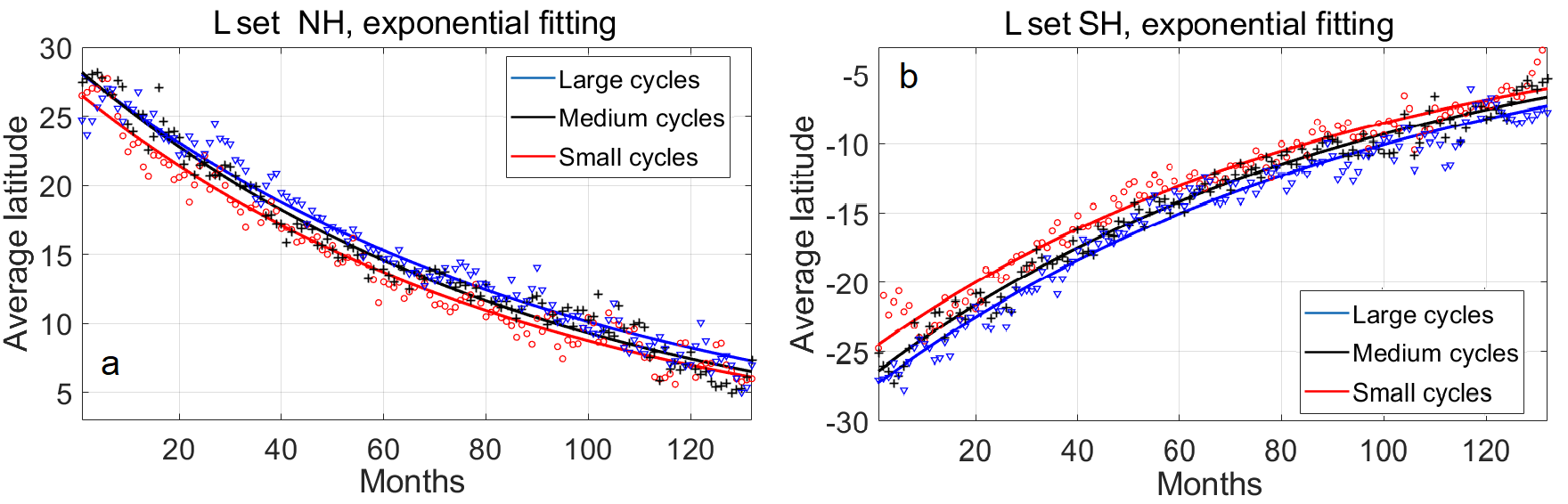}
		\caption{a) The exponential fits of the average drift for large, medium and small cycles of the northern hemisphere. b) Same as a but for the southern hemisphere.}
		\label{fig:L_M_S_exp_fits_Lset}
\end{figure}

We also wanted to know if the strength of the cycle affects to the path of the drift of the SGs. The categories of the large, medium and small cycles are defined similarly to \cite{Hathaway_2011}; large cycles: 18, 19, 21 and 22, medium cycles: 15, 17 20 and 23, small cycles: 12, 13, 14 and 16.  Figure \ref{fig:L_M_S_exp_fits_Lset}a and b show the exponential fits and points of the monthly average latitudes of the L set SGs for the northern and southern hemispheres, respectively. It is evident that the large cycles have higher latitudes and small cycles lowest latitudes through the evolution of the average cycle. This is, however, at least partly due to the fact that the small cycles have on average starting times about seven months earlier than the minimum,
while medium and large cycles have starting times about equal to minimum \citep{Hathaway_2011}. There seems to be, however, a slight difference between the drift paths of the large and medium sized cycles such that the path of the average large cycle is about one degree higher than the average medium cycle, at least at the descending phase of the cycle. This applies for both northern and southern hemisphere of the average cycle. It seems that the drift path of the stronger cycles follow slightly higher latitudes than those of the weaker cycles.

\begin{figure}
	\centering
	\includegraphics[width=1\textwidth]{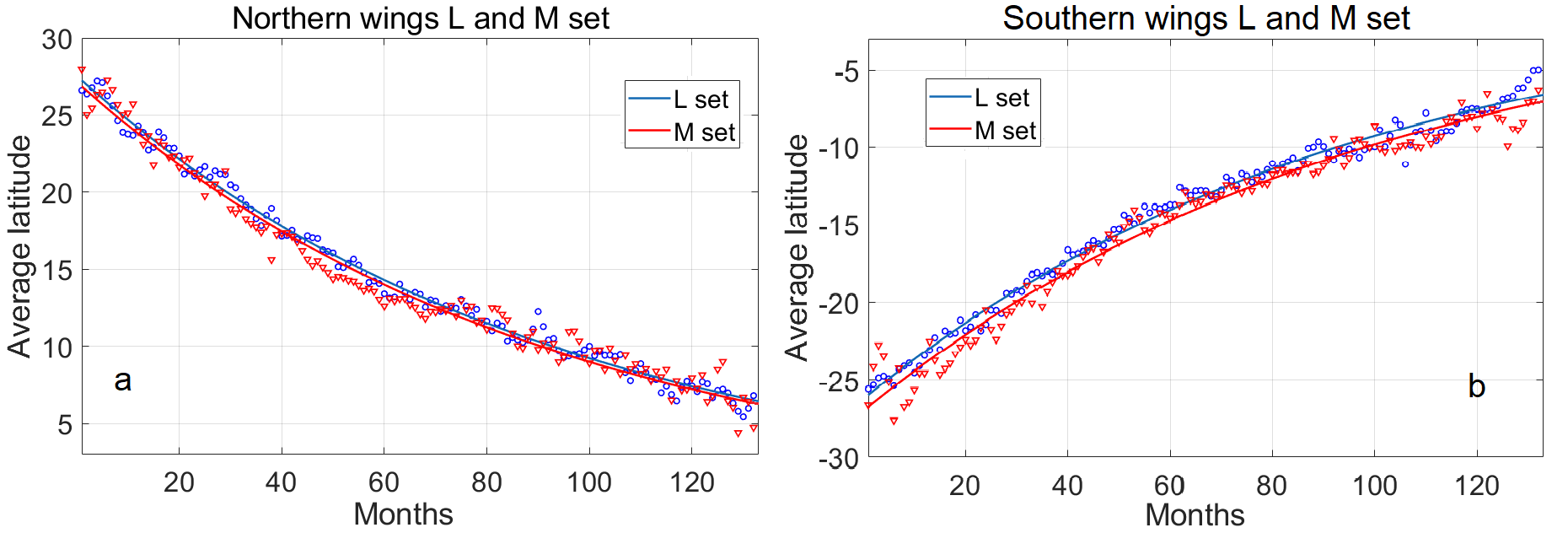}
		\caption{a) The exponential fits of the average drift for the average L and M set cycles of the northern hemisphere. b) Same than a but for the southern hemisphere.}
		\label{fig:L_set_M_set_NH_comparison}
\end{figure}

Another interesting question is, do the areas of the SGs affect to the paths of the groups. While L set has a shortage that it does not contain area information, we must analyse in this respect M set for the Cycles 12\,--\,23. Let us, however, first compare the average NH cycle between L set and M set. Figure \ref{fig:L_set_M_set_NH_comparison}a and b show the L set and M set drift path for the average northern and southern hemispheres, respectively. It seems that the path of the M set is about a half degree higher latitudes than the L set set for the southern hemisphere. (However, when doing an area weighted analysis for the M set, the paths of L set and M set were almost identical). It is evident, that there is no big difference between these datasets, although they differ in the way the the SGs have been recorded, i.e that L set contains only the first appearance of each SG and M set includes daily appearances of the same SGs.

\begin{figure}
	\centering
	\includegraphics[width=1\textwidth]{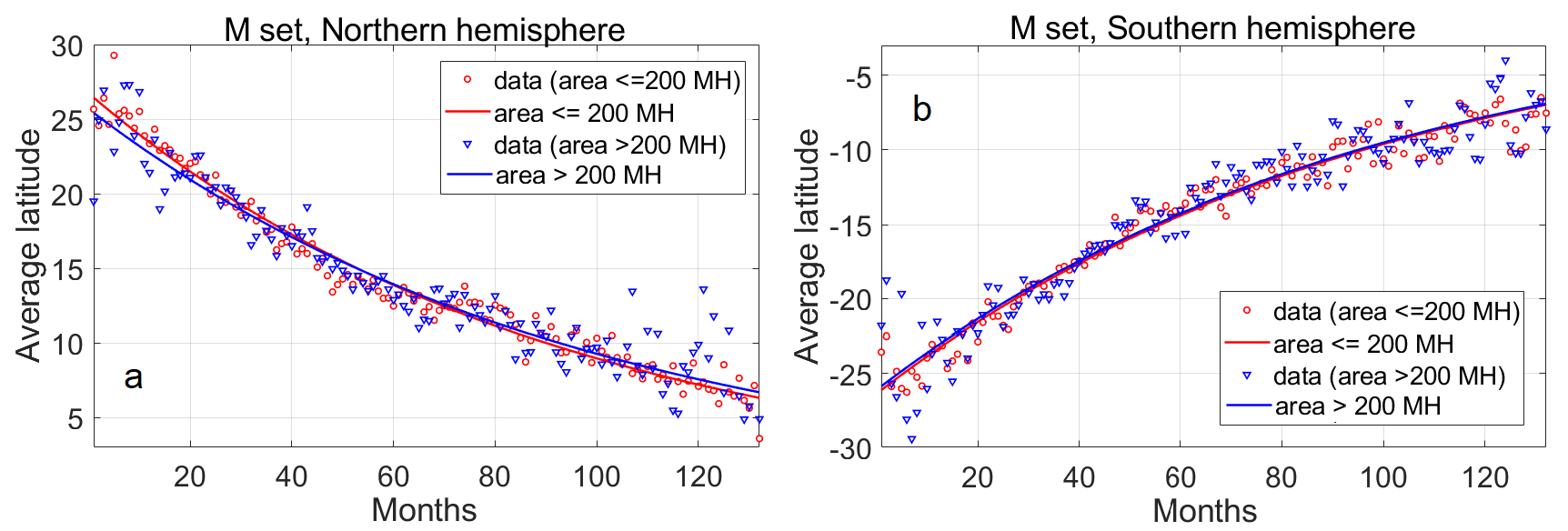}
		\caption{a) The exponential fits of the average drift for small ($<=$200 MH) and large ($>200$ MH) sunspot groups for the northern hemisphere. b) Same as a but for the southern hemisphere.}
		\label{fig:L_set_M_small_large_comparison}
\end{figure}

It is now clear that we can study the drift paths of small and large area sunspot groups using only the M dataset. We divide the SGs into two categories; small SGs with area smaller or equal to 200 MH (millionths of hemisphere), and large SGs with area larger than 200 MH. Figure \ref{fig:L_set_M_small_large_comparison} shows the drift paths for both area categories of the M set for northern hemisphere. We have used here are area weighted calculation of the mean latitude, i.e. $\overline{\lambda}=\sum\,A(\lambda_{i})\lambda_{i})/\sum\,A(\lambda_{i})$. It seems that the area does not affect very much to the drift path of the SGs. There is, however, difference in the northern hemisphere such that small area SGs are located at slightly higher latitudes in the ascending phase and at lower latitudes in the descending phase of the average cycle. This is because the large area SGs are spread mainly between latitudes 10\,--\,25 degrees, but small area SGs more widely between 0\,--\,40 degrees \citep{Takalo_2020_1}. In the southern hemisphere the drift paths are, however, more similar. Note also, that there are quite a lot variations in the data points at the beginning and in the end of the cycles, especially for large category SGs, due to lack of data and thus worse statistics near cycle minima.  

\begin{figure}
	\centering
	\includegraphics[width=1\textwidth]{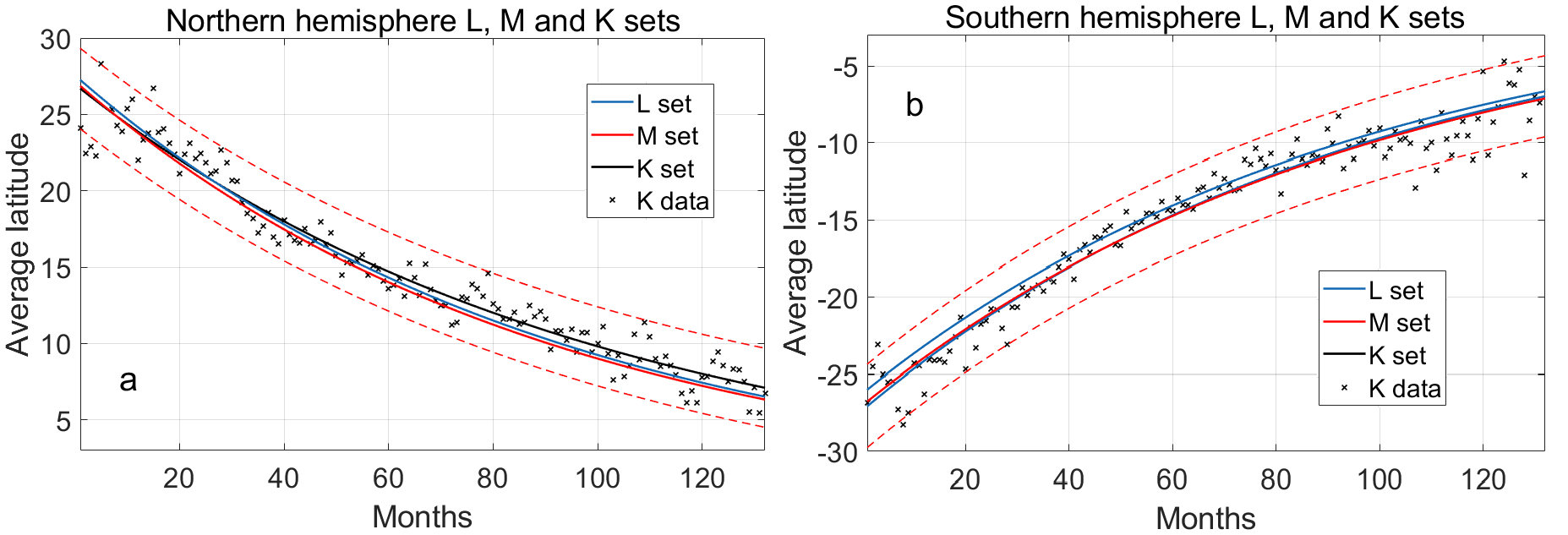}
		\caption{a) The exponential fits of the average drift of L and M sunspot groups and of K set sunspots for the northern hemisphere. b) Same as a but for the southern hemisphere. The data points and 95\,\% confidence limits of K set are also shown in both figures.}
		\label{fig:L_M_K_set_comparison}
\end{figure}

Figure \ref{fig:L_M_K_set_comparison}a and b show the exponential fits of L (C8\,--\,C23), M (C12\,--\,23) and K (C16\,--\,C23) sets with data points of K set for the northern and southern hemispheres, respectively. The points marked with 'x' in the figure and confidence limits (red dashed curves) are those for the K set. Notice the quite wide variation in the beginning and at the end of the average cycle. This is also seen as wider 95\,\% confidence limits than earlier for the L set. These are due to less amount of data compared to L and M datasets. Although the K set (Kodaikanal) data contains separate sunspot data and for a shorter interval of cycles, its seems to scale similarly to the sunspot group datasets. Especially, M set and K set exponential fits are very similar, except for the descending phase of the northern hemisphere.

\section{Conclusion}

We have analysed the monthly sunspot group (SG) data for Solar Cycles 8\,--\,23 and calculated the average latitude of the migration for the northern and southern hemisphere. We find that exponential function fits best to the average drift of the SGs. The R$^{2}$s for the exponential fits are 0.992 and 0.988, but second-order polynomial fits are only slightly worse. The drift velocities are are 0.30 and 0.27 degrees/month (meridional speed$\approx$\,1.41 and 1.27 m/s) in the beginning and 0.071 and 0.069 degrees/month ($\approx$\,0.33 and 0.32 m/s) at the end of the average cycle for northern and southern hemisphere, respectively.
 
We do also bi-linear fits (first-order polynomial fits in two fractions) because they tell us the crossing point (knee) were the drift velocity decreases the most. Furthermore, we find the crossing point such that the difference of the slopes is largest. This crossing point for the Cycles 8\--\,23 is at 50 months and 15.4 degrees of latitude in the northern hemisphere. The crossing point for the southern hemisphere is somewhat further from the start of the cycle, i.e. 54 months and -14.4 degrees of southern latitude. The drift velocities calculated from the slopes of the bi-linear fits are 0.233 and 0.216 degrees/month ($\approx$\,1.10 and 1.02 m/s) before the crossing point and 0.118 and 0.101 degrees/month ($\approx$\,0.55 and 0.47 m/s) after the crossing point. The bi-linear fit is interesting, because the possible abrupt change in the drift velocity and thus in meridional speed near 15 degrees of latitude could be related to Gnevyshev gap. This issue is really worth studying in a forthcoming project.

We also found that there is a slight dependence of the migration on the strength of the cycle such that the drift path of the stronger cycles follow slightly higher latitudes than those of the weaker cycles. The areas of the sunspot groups affect also slightly to the drift path such that the path of large area SGs exists at somewhat lower latitudes in the ascending phase and at higher latitudes in the descending phase of the cycle than those for the smaller area SGs. It is also evident, that the form of the drift path is almost similar for all three dataset (L set, M set and K set), which we have investigated.

\begin{acknowledgements}

The sunspot group data are downloaded from VizieR database (https://vizier.u-strasbg.fr/viz-bin/VizieR?-source=J/A+A/599/ \newline A131\&-to=3). The RGO sunspot area data are from RGO-USAF/NOAA \newline (https://solarscience.msfc.nasa.gov/greenwch.shtml) and Kodaikanal data from https://github.com/bibhuraushan/KoSoDigitalArchive. The dates of cycle maxima were obtained from from the National Geophysical Data Center (NGDC), Boulder, Colorado, USA (ftp.ngdc.noaa.gov). I am grateful to J.-G. Richard for useful communication.

\end{acknowledgements}

\textbf{Disclosure of Potential Conflicts of Interest}

The author declares that there are no conflicts of interest.

%
%
\bibliographystyle{spr-mp-sola}
\bibliography{references_JT_SolPhys}  
%
%
%
%

\end{article} 
\end{document}